\documentclass[12pt,]{article}
\DeclareUnicodeCharacter{2032}{\ensuremath{\prime}}

\usepackage{authblk}
\usepackage{fullpage}
\usepackage{blkarray}
\usepackage{amssymb,amsmath}
\usepackage[utf8x]{inputenc}
\usepackage[T1]{fontenc}
\usepackage{siunitx}
\usepackage[version=3]{mhchem}
\usepackage{subfiles}
\usepackage[numbib]{tocbibind}
\usepackage[nonumberlist]{glossaries}  
\usepackage{tcolorbox}

\usepackage[numbers,comma,sort&compress]  
   {natbib}

\makeglossaries
\usepackage{setspace}
\usepackage[unicode=true]{hyperref}
\hypersetup{breaklinks=true,
            bookmarks=true,
            colorlinks=false,
            pdfborder={0 0 0}}
\usepackage{graphicx}
\graphicspath{{figures/}}
\makeatletter
\def\ScaleWidthIfNeeded{%
 \ifdim\Gin@nat@width>\linewidth
    \linewidth
  \else
    \Gin@nat@width
  \fi
}
\def\ScaleHeightIfNeeded{%
  \ifdim\Gin@nat@height>0.9\textheight
    0.9\textheight
  \else
    \Gin@nat@width
  \fi
}
\makeatother
\setkeys{Gin}{width=\ScaleWidthIfNeeded,height=\ScaleHeightIfNeeded,keepaspectratio}%

\title{Can genomic analysis actually estimate past population size?}
\author[1,*]{Janeesh K. Bansal}
\author[1]{Richard A. Nichols}
\affil[1]{\footnotesize School of Biological and Behavioural Sciences, Queen Mary University of London, London UK}
\affil[*]{\footnotesize Correspondence: j.k.bansal@qmul.ac.uk (J.K. Bansal)}

\date{\today}

\begin{document}

\linespread{1.5}

\maketitle


\section{Keywords}

\begin{itemize}
    \item Demographic inference
    \item Sequentially Markovian Coalescent 
    \item Bottleneck
    \item Population Structure
    \item Range contraction and expansion
\end{itemize}


\section{Abstract}


Genomic data can be used to reconstruct population size over thousands of generations, using a new class of algorithms (SMC methods). These analyses often show a recent decline in $N_e$ (effective size), which at face value implies a conservation or demographic crisis: a population crash and loss of genetic diversity. This interpretation is frequently mistaken. Here we outline how SMC methods work, why they generate this misleading signal, and suggest simple approaches for exploiting the rich information produced by these algorithms. In most species, genomic patterns reflect major changes in the species' range and subdivision over tens or hundreds of thousands of years. Consequently, collaboration between geneticists, palaeoecologists, palaeoclimatologists, and geologists is crucial for evaluating the outputs of SMC algorithms.

\newpage

\section{Problems in inferring past population size}

Genomes carry information about a species’ evolutionary history. Consider the higher genetic diversity found in humans from African populations compared to the rest of the world. This pattern has been explained by genetic bottlenecks during the Early Modern Human migrations out of Africa. The onset of migration can be dated, using genetic evidence, as starting between 50,000 and 100,000 years ago, depending on the assumptions of the analysis and the samples used \cite{lopez2015human}. Those events are particular to humans but, since then, almost all species will have undergone comparable dramatic fluctuations in population size and species range. These more recent events were caused by the changes in global climate that culminated in the last glacial maximum 20,000 years ago (at high latitudes), and the subsequent warming \cite{hewitt2004genetic}. These demographic changes will have shaped the patterns of genetic diversity in each species; for example, an analysis of wild boar genomes detects the signatures of population decline in both Europe and China, as the glacial maximum arrived \cite{groenen2012analyses}. Similar patterns are widespread and may even extend to marine species \cite{vijay2018population}.

Demographic information on prehistoric events can be obtained from extraordinarily small numbers of genomes. Ingenious innovations in genomic analysis allow the reconstruction of the history of effective population size even from a single diploid individual \cite{li2011inference, schiffels2014inferring}. An example is shown in Figure 1 (Key Figure), in which the curves show an ancestral population size of 150,000 which has declined in the most recent 200,000 generations. Biologists have embraced this type of analysis with enthusiasm. Worryingly, a series of papers have recently pointed out \cite{mazet2015demographic, rodriguez2018iicr, arredondo2021inferring} that this type of pattern may not represent a population decline at all. Figure 1 illustrates this problem. Both graphs were generated by the analysis of genomic data generated by simulations; we therefore know the true demographic history. Only Figure 1A shows a population that has actually declined in size. In Figure 1B the population size was constant, but the population was subdivided.

This is not a simple case of effective population size, $N_e$, being different from the census size; an idea which is familiar to most geneticists from textbook introductions to $N_e$ (Box 1). We argue here that this false signature of population decline, due to population subdivision, will be present in most species, and that there are additional equally important sources of misleading estimates of $N_e$. Hence, the short answer to the question `Do these genomic analysis tools actually estimate the past population size?' is no. On a more positive note, we propose some simple principles that biologists can use to interpret these trends to obtain insights into the populations’ demography, history and ecology.


\section{How modern methods work}

It might seem implausible that the analysis of a single genome can provide such rich information about the history of a population. After all, that is a sample of only two haploid genomes: the maternal and paternal sets of chromosomes. Such a meagre sample might appear far too small for reliable inference. However, if we trace the ancestry of the genome backwards in time, it becomes apparent that the sample size can be thought of as much larger. Two generations back most humans have four grand-parents, and only six generations back we could each have as many as $2^6 = 64$ ancestors. Although we only have 46 chromosomes, more ancestors from this generation could have contributed to our genome, because of the effects of recombination. Each chromosome is typically composed of segments from different ancestries, brought together by recombination in the parents’ germ line (approximately one chiasma per chromosome arm each generation \cite{de2001recombination}). The genetic patterns in human genomes are explained by much deeper ancestries, tracing back over a hundred thousand generations and hence the lineages have passed through millions of ancestors distributed over that time. The stitching together of segments by recombination down all these generations culminates in the patchwork making up the genomes in our sample.  

The inference of past demography exploits these patterns by comparing haploid genomes. We can imagine tracing the ancestry of a particular genomic region backwards in time along the maternal and paternal lineages until we find a shared segment of the genome in a common ancestor. This coming together is known as a coalescent event. The coalesced block has left two descendant segments (one maternal and one paternal) in our sampled genome. A pair of segments descended from a more recent ancestor will on average be longer and have fewer differences between the maternal and paternal genomes, because there have been fewer generations of recombination and mutation. Unfortunately, we cannot look at a genome and unambiguously identify the junctions between different segments, but innovative software overcomes this limitation. Figure 2 illustrates the approach applied to the analysis of simulated data. Figure 2A has zoomed in on 1000 bp of the genome. Because the data come from a simulation, we know the location of junctions between segments and times of ancestry, which are shown by a dashed line. We can see that the older segments tend to be shorter and have more differences between the two genomes (shown by the vertical lines). The black points show the coalescence times inferred by the algorithm, which exploits the genetic differences to try and reconstruct the patchwork of segments. The time estimate for a particular location on the genome is an average over different iterations of the algorithm, each of which proposes the positions of junctions between the segments (these positions will often differ between iterations, as the evidence for their precise location is typically weak).  

The timing of coalescent events can be related to past population size. If there were an era when the population size was small, we might expect to see an excess of coalescence times during that period, because the limited number of possible ancestors would lead to more coalescent events (see Box 2). The converse would be true for a period of large population size. From Figure 2A we can see that the inferred timings for any one region of the genome are only approximately related to the true values. However, this relatively unreliable information can be combined across the genome to build up a more accurate picture of the coalescence rate, $r$, at different times in the past and hence the effective population size $N_e \propto \frac{1}{2r}$ (Figure 2B).

Ideally, we would want an algorithm to generate a sample of possible ancestral relationships in the form of \textbf{ancestral recombination graphs} (See Glossary) \cite{lewanski2023era} (ARGs) according to the probabilities calculated from the differences between the genomes. However, this problem is computationally demanding and often intractable \cite{wilton2015smc}. Instead, there has been a flourishing of algorithms making use of simplifying assumptions that allow a sample of ancestries to be drawn from \textbf{hidden Markov models (HMMs)} \cite{mcvean2005approximating} giving a very close approximation to the same distribution. In essence, an algorithm moves along the genome putting in break points between different segments with different ancestral origins and times (the hidden states). 

They are based on the \textbf{sequentially Markovian coalescent (SMC)} \cite{mather2020practical}, which has been refined over different software releases to allow for the consequences of coalescent events for adjacent segments and to exploit efficient approximations of the coalescence time distribution (the development from \textbf{PSMC} to SMC’ and Gamma-SMC  \cite{li2011inference, marjoram2006fast, schweiger2023ultra}). Developments for exploiting larger sample sizes (\textbf{MSMC}, \textbf{MSMC2}) allow the analysis of the more recent past: in humans increasing the sample size from one to eight allows the examination of events 2,000 years ago, instead of 20,000 years \cite{schiffels2020msmc}.


\section{Changing $N_e$ estimates in a constant population}

In Figure 1B we saw an example of the analysis of genomic data from a population of constant size: yet the $N_e$ estimates changed with time. For the recent past the estimate was much smaller than the population size and for the more distant past it was an overestimate. The population in question was subdivided, consisting of partially isolated demes connected by gene flow. These effects are best understood by considering the rate of coalescence. In the recent past there will have been rapid coalescences between lineages in the same deme, hence it appears that $N_e$ was small and of the same order as the deme size. As we look further back in time these opportunities for rapid coalescence become exhausted and some ancestral lineages will have made their way into separate demes, and so escape coalescence for many generations, giving the appearance of an inflated $N_e$. Wakeley called this change in coalescent rate a transition between the ‘scattering’ and ‘collecting’ regimes, which he modelled to explain the puzzling allele frequency spectrum in human data \cite{wakeley1999nonequilibrium}. More recently the field has moved on from the analysis of allele frequency data to the interpretation of whole-genome data, and the use of SMC-based methods. Because these methods infer the change in coalescence rate over time, they show this transition explicitly (for example the decline in coalescence rate shown in Figure 2B) although the value plotted is usually $N_e$ ($ \propto \frac{1}{2r}$). The labelling of the Y axis as $N_e$ in plots of SMC output (such as Figure 1) is perhaps unhelpful, except for cases which can be approximated by a model of a single random breeding population. When Mazet $et \, al$  \cite{mazet2016importance} pointed out that the outputs of SMC-type analyses were being wrongly interpreted as reflecting changes in population size, they proposed the term IICR (Inverse Instantaneous Coalescence Rate) as an alternative, to remind us to avoid these errors of reasoning. 

In order to interpret the patterns in these plots, we need models of population subdivision which will output the timings and the magnitude of the change in coalescence rates. Figure 3 illustrates two tractable models. In the island model (first row) there are a number of demes (five in this example) which exchange migrants with each other. As we trace the ancestry backwards in time some proportion of the genome will coalesce (move to state \textbf{C}) where it will stay (coalescence cannot be undone).  Tracing ancestry backwards in time some lineages will move apart, due to different histories of migration, hence the pair transitions from state \textbf{S} (same deme) to state \textbf{D} (being in different demes). These parts of the genome will take much longer to coalesce as another migration event is needed to move the two lineages into the same deme (\textbf{S}) before they can coalesce.

These dynamics can be captured by a transition matrix (see Box 3), which gives the expected proportion of the genome that moves between or within states each generation (\textbf{S} $\rightarrow$ \textbf{S}, \textbf{S} $\rightarrow$ \textbf{D}, \textbf{D} $\rightarrow$ \textbf{S} and \textbf{D} $\rightarrow$ \textbf{D}). The coalescence rates generated by this calculation (converted to IICR or $N_e$) are shown as black curves in the right-hand column of Figure 3. They closely match the values estimated by MSMC2 from simulated data generated by the same model (coloured curves). As an alternative to iterating the matrix multiplication, it is possible to use eigen analysis of the matrix (explained in \cite{seabrook2023tutorial}) to calculate the same black curve.

The same principles for finding the expected trend in IICR ($N_e$) can be applied to a stepping stone model of population subdivision, in which demes only exchange migrants with adjacent populations. There is a minor increase in complexity (we need to specify states that keep track of the displacement between the two lineages in the X and Y directions), so there is a correspondingly large transition matrix; but otherwise, the principles are the same. Examples of different displacements are shown by labels \{$x, y, u, v$\} in Figure 3 . Where x $\leftrightarrow$ y are in the same deme (0 in the 2nd column), x $\leftrightarrow$ u are in different demes separated by a single horizontal step (1) and x $\leftrightarrow$ v are separated by one horizontal step and one vertical step (2). These potential displacements are the states of the stepping stone model. 

We can learn some important lessons from these curves. For example, comparisons between the last two columns of Figure 3 show that increasing migration rate actually decreases the inferred $N_e$, a principle which Wakeley suggested might apply to human history:  $N_e$ could have decreased as our ancestors became more mobile \cite{wakeley1999nonequilibrium}.  Many geneticists find the association of greater mobility with smaller $N_e$ (and greater isolation with larger $N_e$) counter-intuitive. It is perhaps better to think in terms of coalescence rates: if populations exchange more migrants, then lineages in different demes are more likely to be able to come together and coalesce.


\section{Using our biological insight}

Biologists will typically have sufficient knowledge of their study organism to decide if the population should be considered subdivided. From our understanding of animal movements and the dispersal of pollen, seed and other propagules in plants (and other kingdoms) it is possible to propose a population structure. We therefore recommend that the interpretation of how the IICR ($N_e$) changes over time should proceed hand in hand with calculation of the expected trend; that could be generated by a model of the form shown in Figure 3, some other analytical model or a simulation (e.g. using SLiM \cite{haller2019slim}). It need not be a definitive model, but the exploration of the patterns from a choice of plausible models is fundamental to evaluating explanations for the observed trends. Allowing for population structure can yield qualitatively different interpretations of genomic data, including the suggestion that there was little interbreeding between early modern humans and Neanderthals \cite{chikhi2018iicr, tournebize2025ignoring}. Secondly, the time scale of most IICR analyses is of the order of hundreds of thousands of years. The interpretation therefore needs to take into account what we know about the palaeoecology of our study organism. Only 20,000 years ago, most species distributions were completely different from today. Figure 4 shows a reconstruction of the distribution of North European trees at the last glacial maximum \cite{birks2008alpines}. They are fragmented in an entirely different way from the current population structure. Of course the associated species will be similarly effected too. The arrows show the possible directions of the initial range expansions as the climate warmed. Counter-intuitively, these population expansions would be expected to lead to declines in $N_e$ as a limited number of individuals at the expanding margin of the species range contribute to the establishment of the species in the newly available territory \cite{ibrahim1996spatial, hewitt2005genetic, excoffier2008surfing}, in a phenomenon sometimes called gene surfing.  Hence periods of $N_e$ decline may coincide with improving conditions and population expansion. Finally, there may be evidence of such episodes occurring repeatedly since the climate has cycled with a 100,000 year period throughout the Quaternary period \cite{hewitt2004genetic}.  

Using our knowledge about population structure, it is possible to carry out more incisive analysis. It is particularly informative to make comparison between genomes known to come from different subpopulations, or groups having more ancient histories of separation. In the case of humans, we can draw on historical, archaeological, cultural and linguistic evidence to identify the groups of interest, an approach Song $et\ al$ applied \cite{song2017modeling} to African populations. They matched the PSMC output from these comparisons to the results of the same analysis applied to genomic data generated by simulations. Of particular interest was the timing of splits among their study populations (including Yoruba, Esan, Maasai, Mende and populations outside Africa). Rather than informally comparing the curves, they used \textbf{Approximate Bayesian Computation} (ABC) to estimate the range of split-times consistent with the data, a refinement which should be more widely used for the study of other species. These reconstructions of past demography are not only of interest in themselves, but also as a starting point for further analyses, particularly the detection of selection. A joint analysis of ancient demography and selection is possible \cite{hsieh2016whole} and should avoid misattributing the effects of random genetic drift to selection. Recent innovations can use much larger samples of genomic sequence, which will soon be commonplace \cite{huang2025estimating}. They use algorithms which can encode genomic data from these large samples by inferring features of the ARG (the succinct tree sequence, TS), and then exploit features that can be derived from the TS to generate statistics that are, in turn, fed into an ABC (See outstanding questions).


\section{Tapping new expertise}

We have emphasised the importance of taking into account the time scale of the IICR reconstructions generated by the analysis of genomic data. In most species, including our own, the patterns we observe in today's genomes will have been generated by ancient events, some occurring during times when our study species were living in distant locations and different climates to the present day. Our genetical and biological training has not typically equipped us to confidently reconstruct this history. We therefore need to draw on the expertise of colleagues to help. Experts in the analysis of pollen and Foraminifera deposits can reconstruct the palaeoecology.  Archaeologists and palaeontologists tend to be much more familiar with the relevant discoveries. Geologists and palaeoclimatologists can provide us with reconstructions of climate oscillations and the timing and speed of other superimposed changes. These collaborations can be exciting and fruitful, and sometimes our genetic evidence provides important new information too.


\section{Acknowledgments}

We thank Dr Robert Verity, Dr Matteo Fumagalli, Nichols Lab, Fumagalli Lab and two anonymous referees for helpful suggestions and discussions. We acknowledge the assistance of the ITS Research team at Queen Mary University of London, and ITS-R's support of QMUL's Apocrita HPC facility (http://doi.org/10.5281/zenodo.438045). JKB is supported by a QMUL Principal's Studentship.  


\section{Declaration of Interests}

No conflicts of interest to declare.



\bibliography{ref_final}


\section{Figure legends}


\subsection{Figure 1}

\begin{figure}[hbt!]
    \centering
    \includegraphics[width=1\linewidth]{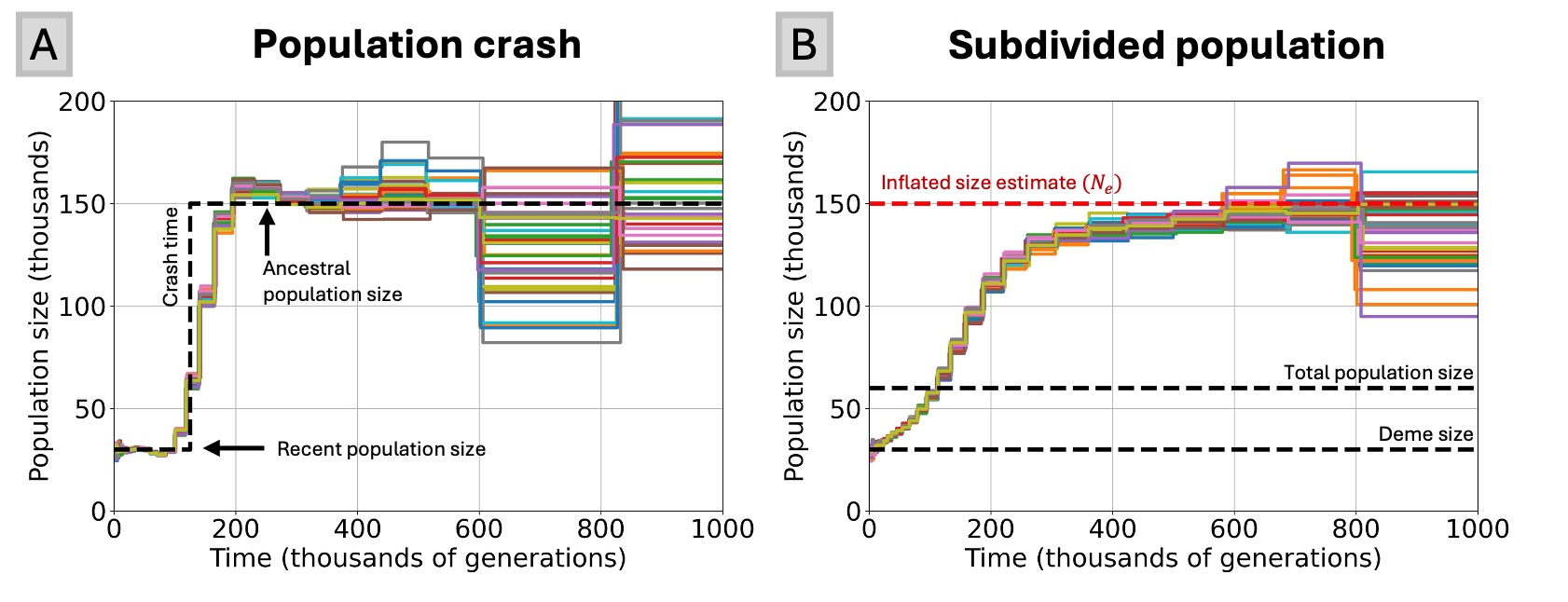}
\end{figure}

\begin{itemize}
    \item \textbf{Title:} \textbf{Population decline or population substructure?}
    \item \textbf{Main:} Estimates of effective population size over the last million generations obtained by applying the MSMC2 algorithm to simulated data. Coloured curves show the output from $50$ simulations. The black dashed lines shows the true parameters for the models that generated the simulated data. 
    [A] A single population. The population size dropped instantaneously from 150,000 to 30,000 125,000 generations ago.
    [B] The population is divided into two demes of size 30,000 connected by migration at rate $2 \times 10^{-6}$ per generation. The population size is constant. The red dashed line shows the inflated size estimated by the algorithm. 
    Notice that both cases appear to show a recent decline in population size, but that is only true in the first case.
\end{itemize}

\subsection{Figure 2}

\begin{figure}[hbt!]
    \centering
    \includegraphics[width=1\linewidth]{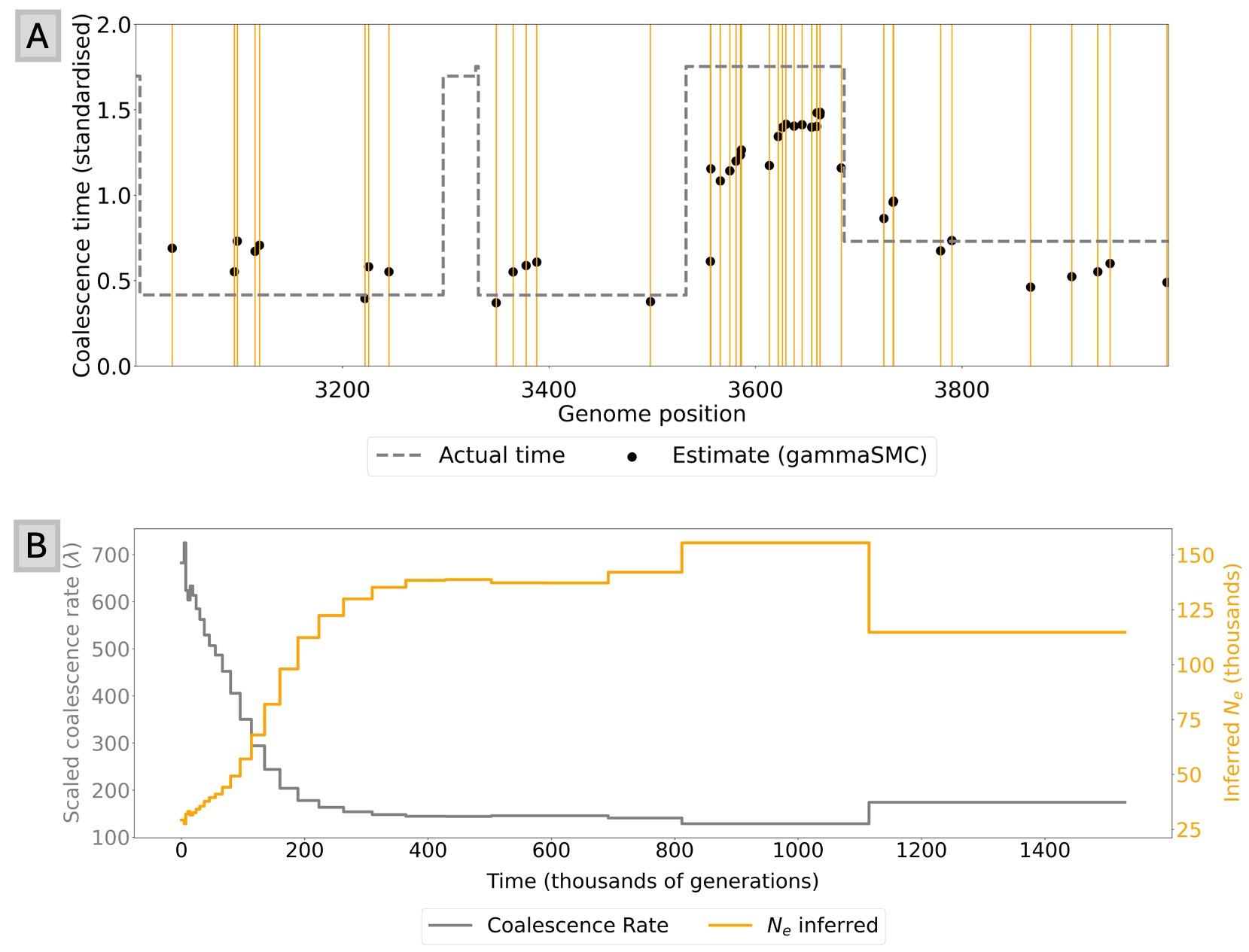}
\end{figure}

\begin{itemize}
    \item \textbf{Title:} \textbf{Analysis of the differences between two haploid genomes}
    \item \textbf{Main:} [A] A section of a simulated genome. The vertical orange lines represent the position of genetic differences between the two genomes. The dashed line gives the true time back to the common ancestor. The black dots show the estimates of time back to a common ancestor at different locations along the genome estimated by Gamma-SMC. Notice that they are correlated with the true values but not very precisely. [B] The estimate of the coalescence rate $(\lambda)$ at different times before the present obtained by MSMC2 from the whole genome (dark curve) in an island model . The same data can be displayed as $N_e$ values (orange curve), which might be better termed Inverse Instantaneous Coalescent Rates.  
\end{itemize}

\subsection{Figure 3}

\begin{figure}[hbt!]
    \centering
    \includegraphics[width=1\linewidth]{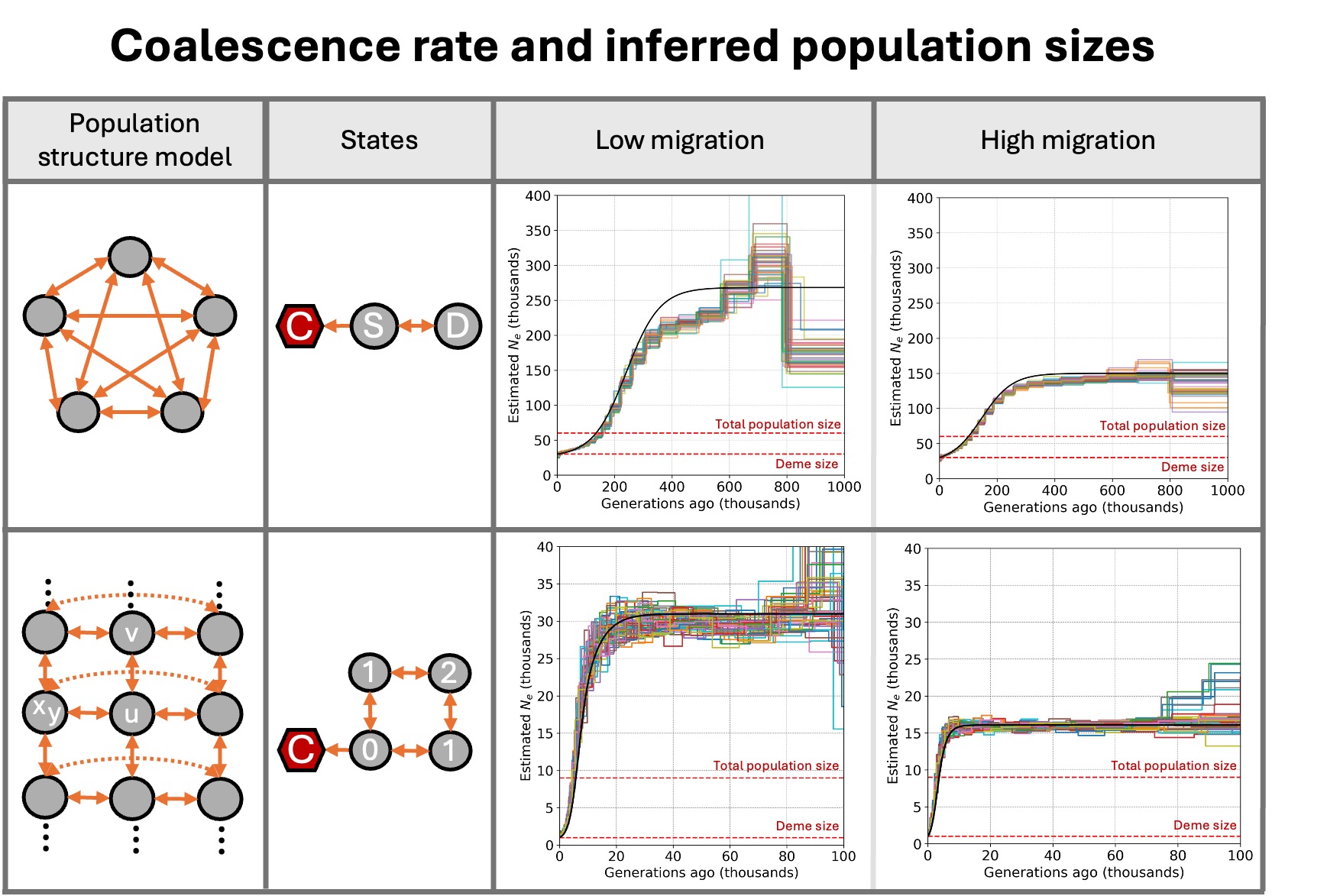}
\end{figure}

\begin{itemize}
    \item \textbf{Title:} \textbf{The signatures of population substructure}
    \item \textbf{Main:} The first row illustrates an island model with $5$ demes and constant migration between every pair.  The second row illustrates a stepping stone model (3x3 demes on a torus). The first column displays these arrangement of populations. The second column shows the different states that a pair of lineages could be in as their ancestry is traced back through preceding generations. In the island model, there are only three different states: they could be in different demes (\textbf{D}), the same deme (\textbf{S}) or in the absorbing state of being coalesced (\textbf{C}). In the stepping stone model being coalesced (\textbf{C}) is again an absorbing state, the lineages could be in the same deme (\textbf{S} or \textbf{0} in this model), but there are now several different states when the pair of lineages are in different demes, depending on the displacement between them. The labels \{$x,y,u,v$\} show possible locations of lineages, with different displacements. The final two columns show the results from an MSMC2 analysis of simulated genetic data produced by these models (coloured curves) and the expected trends (black curves), for low and high migration rates.  For low migration case, the parameter values were: island model, $Nm=0.027$; stepping stone model, $Nm=0.025$. For high migration case, the parameter values were: island model, $Nm=0.058$; stepping stone model, $Nm=0.075$. Notice that the lower height of the curves show that for both models lower migration rates were associated with higher $N_e$ estimates.
\end{itemize}

\subsection{Figure 4}

\begin{figure}[hbt!]
    \centering
    \includegraphics[width=1\linewidth]{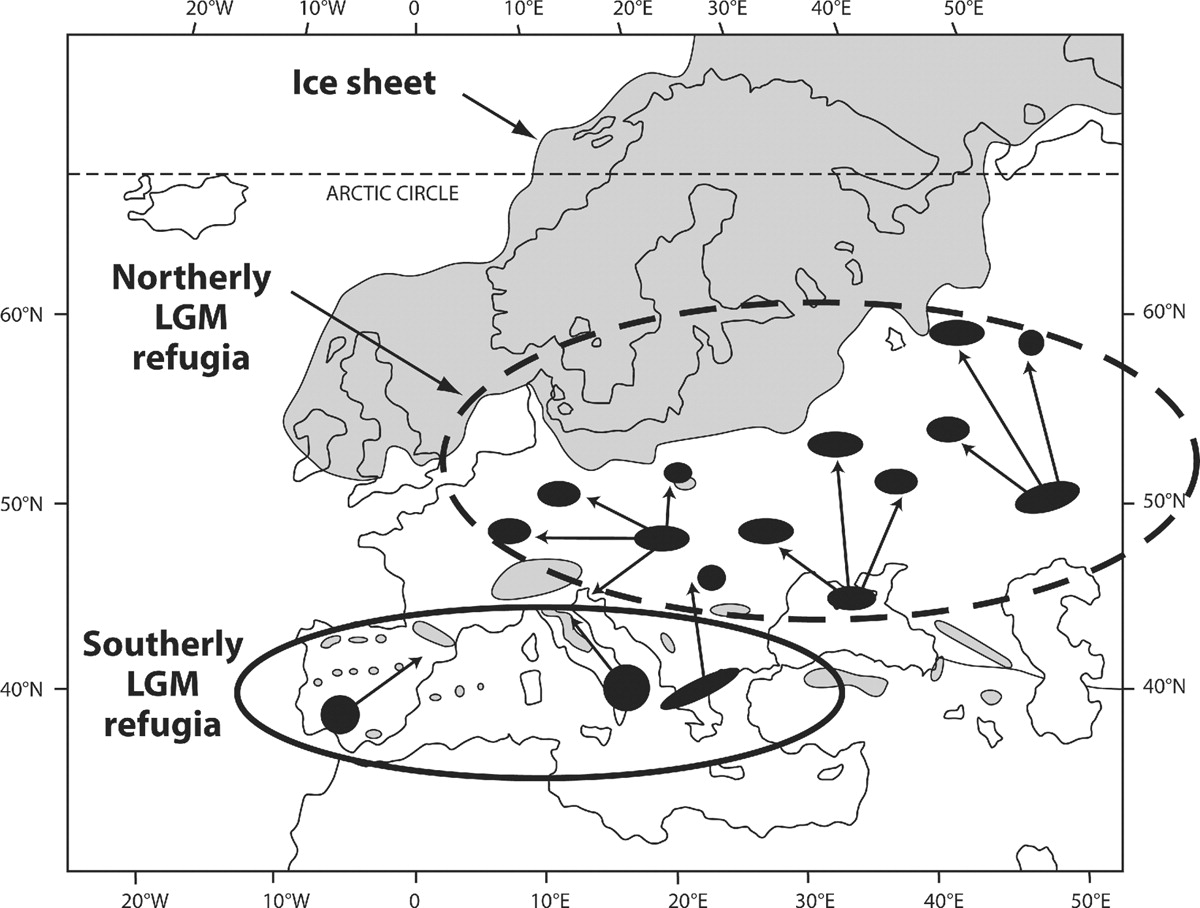}
\end{figure}

\begin{itemize}
    \item \textbf{Title:} \textbf{Range expansion after the LGM}
    \item \textbf{Main:} Figure taken from \cite{birks2008alpines} proposing the location of refugia for trees during the last glacial maximum, with arrows showing the direction of Holocene postglacial range expansion (reproduced with permission).
\end{itemize}


\glsaddall

\section{Glossary}

\newglossaryentry{Ancestral Recombination Graph (ARG)} {name=Ancestral Recombination Graph (ARG), description={: A graph representing the genealogical relationships among a set of sampled sequences. Each locus in the genome is associated with a bifurcating tree (subgraph of the ARG) in which the nodes represent coalescent events and the edges show the links between ancestors and descendants. Adjacent loci will tend to share the same tree, but at some point in the past their ancestries can become separated due to recombination. These separation events are represented by branching points on the graph.  When there are only two haploid samples the ARG is very simple, comprising of subtrees with a single node and two descendant branches with a length corresponding to the time to a common ancestor}}

\newglossaryentry{Approximate Bayesian Computation (ABC)} {name=Approximate Bayesian Computation (ABC), description={: A computational method used to perform Bayesian inference when the likelihood function is difficult or impossible to compute. It compares summary statistics obtained from the analysis of the real data with same statistics obtained from data generated by simulations. The distances between the simulated statistics and real statistics are used to infer the range of plausible values  for the parameters of the model (the posterior distribution).  One possible choice of statistics is the values of $N_e$ estimated at different dates by an SMC algorithm}}

\newglossaryentry{Hidden Markov Model (HMM)} {name=Hidden Markov Model (HMM), description={: An HMM analyses observable values whose outcomes depend on an underlying process which cannot be observed directly.  The modelling is made easier by the assumption that it is a Markov process. In other words, whilst the current state (for example the time to ancestry at a location on the genome) may be required to model the next value, the process is memoryless: it does not depend on preceding states}} 

\newglossaryentry{Multiple Sequentially Markovian Coalescent (MSMC and MSMC2)} {name=Multiple Sequentially Markovian Coalescent (MSMC and MSMC2), description={: An algorithm that uses the SMC' approach to infer the demography of a population using multiple samples. MSMC retains simplicity by modelling the first coalescent event among the sampled genomes, whereas MSMC2 describes the entire distribution of pairwise coalescence times.}}

\newglossaryentry{Pairwise Sequentially Markovian Coalescent (PSMC)} {name=Pairwise Sequentially Markovian Coalescent (PSMC), description={: An algorithm which uses the SMC approach to infer the demography of a population using two haploid samples (e.g. one diploid genome)}}

\newglossaryentry{Sequentially Markovian Coalescent (SMC)} {name=Sequentially Markovian Coalescent (SMC), description={: A framework to approximate the ARG. It treats ancestry as Markovian between adjacent loci along a chromosome. This means that the ancestry at one locus depends only on that inferred for the previous locus. It includes the effects of recombination events, which change the ancestry at adjacent loci: the junctions between segments of the genome sharing an ancestral coalescent event}}

\renewcommand*{\glossaryname}{}
\printglossaries


\section{Textboxes}

\newtcolorbox{boxA}{
    boxrule = 0pt, 
    title = Box 1: What is $N_e$ ?
}
\begin{boxA}

Small populations will lose genetic diversity through random genetic drift. This process can be measured by the rate of loss of heterozygosity. In an idealized random breeding population of size $N$, the loss occurs at the rate $\frac{1}{2N}$ per generation. If we have measurements of the loss of heterozygosity in a study population, we can therefore calculate the corresponding \textbf{effective population size} $N_e$. \vspace{0.2cm}

The effective population size, calculated in this way, is often smaller than the number of breeding adults in a study population. Population genetics textbooks point out many different reasons why we should expect this discrepancy. For example, in the real population, a minority of breeding individuals may contribution disproportionately to the next generation because they have larger families, monopolize matings or belong to the rarer sex \cite{charlesworth2010elements}.  Nunney has estimated that these effects would be expected to reduce $N_e$ to around half \cite{nunney1991influence}, whereas empirical estimates suggest that $N_e$ is often of the order of $\frac{1}{10}$th of the census size \cite{frankham1995effective}. The additional reduction is most likely due to the effect of fluctuations in population size over the generations \cite{vucetich1997fluctuating}. This lower $N_e$ indicates that genetic diversity is being lost faster that would be expected from the size of the study population. On the other hand, the loss of heterozygosity in a population subdived into demes can be slower than the census size (the total size of all demes) \cite{hossjer2015eigenvalue}. This effect can be explained by the models illustrated in Figure 3.
\end{boxA}

\newtcolorbox{boxB}{
    boxrule = 0pt, 
    title = Box 2: $N_e$ and coalescence rate

}
\begin{boxB}

The value of $N_e$ is inversely related to the rate of coalescence. The broad principle is familiar from our knowledge of relatedness in humans.  In a small isolated village we would expect individuals to share recent common ancestors, so as we trace ancestry back we would come to a coalescent event relatively quickly. The converse applies in a large city. To calculate the rate of coalescence in an idealized random breeding population, we can imagine tracing the ancestry of two autosomal segments of the genome back a single generation. There are $2N_e$ copies of the relevant chromosome in the previous generation, each equally likely to carry the ancestral segment (twice the number of diploid individuals). The chance of both lineages deriving from the same segment is therefore $\frac{1}{2N_e}$

\end{boxB}

\newtcolorbox{boxC}{
    boxrule = 0pt, 
    title = Box 3: Transition matrices capture the dynamics in subdivided populations
}
\begin{boxC}
    
For the island model, (see Figure 3), a 2 by 2 matrix describes the change in the proportion of the genome in states \textbf{D} and \textbf{S} each generation as we trace its ancestry back (to give proportions $D_t$ and $S_t$, after $t$ generations). The first row of the matrix concerns the state \textbf{S}. Segments of the genome are lost from this state due to coalescence, occurring at a rate of $\frac{1}{2N}$ and migration of a lineage out of the deme at rate $2M$ (giving us the term $1 - \frac{1}{2N} - 2M$).  Segments enter state $S$ if migration brings lineages into the same deme ($\frac{2M}{d-1}$) where $d$ is the number of demes. The second row gives the gains to state \textbf{D} due to migration out of state \textbf{S} ($2M$) and the losses when migration brings two lineages together in the same deme ($1-\frac{2M}{d-1}$). 

\begin{gather}
\begin{bmatrix}
    S_{t+1} \\
    D_{t+1}
\end{bmatrix}
=
\begin{bmatrix}
    1 - \frac{1}{2N} - 2M & \frac{2M}{d-1} \\
    2M & 1 - \frac{2M}{d-1}
\end{bmatrix}
.
\begin{bmatrix}
    S_t \\
    D_t
\end{bmatrix}
\label{island_model_trans_mat}
\end{gather}

The initial conditions when sampling both lineages from the same deme are:

\begin{gather}
\begin{bmatrix}
    S_{t=0} \\
    D_{t=0}
\end{bmatrix}
=
\begin{bmatrix}
    1 \\
    0
\end{bmatrix}
\end{gather}

We can calculate the proportion of the genome expected to be in each state each generation back, by iteratively applying equation (1). The proportion of the genome coalesced, $C_t$, can then be calculated as $C_t = 1 - S_t - D_t$ and the coalescence rate as $r_t = C_t - C_{t-1}$.  

\end{boxC}


\end{document}